\documentclass[aps,prb,showpacs,unsortedaddress,twocolumn]{revtex4}
\usepackage{graphics,bm}
\usepackage{epsfig,epsf}
\usepackage{graphicx,amsfonts,amsbsy,amssymb}

\begin{document}

\title{Effective temperature and Gilbert damping of a current-driven localized spin}

\author{Alvaro S. N\'u\~nez}
\email{alvaro.nunez@ucv.cl}
\homepage{http://www.ph.utexas.edu/~alnunez}

\affiliation{Departamento de F\'isica, Facultad de Ciencias
Fisicas y Matematicas, Universidad de Chile, Casilla 487-3, Codigo
postal 837-0415, Santiago, Chile}

\author{R.A. Duine}
\email{duine@phys.uu.nl} \homepage{http://www.phys.uu.nl/~duine}

\affiliation{Institute for Theoretical Physics, Utrecht
University, Leuvenlaan 4, 3584 CE Utrecht, The Netherlands}

\date{\today}

\begin{abstract}
Starting from a model that consists of a semiclassical spin
coupled to two leads we present a microscopic derivation of the
Langevin equation for the direction of the spin. For
slowly-changing direction it takes on the form of the stochastic
Landau-Lifschitz-Gilbert equation. We give expressions for the
Gilbert damping parameter and the strength of the fluctuations,
including their bias-voltage dependence. At nonzero bias-voltage
the fluctuations and damping are not related by the
fluctuation-dissipation theorem. We find, however, that in the
low-frequency limit it is possible to introduce a
voltage-dependent effective temperature that characterizes the
fluctuations in the direction of the spin, and its
transport-steady-state probability distribution function.
\end{abstract}
\vskip2pc

\maketitle

\def\bx{{\bf x}}
\def\bk{{\bf k}}
\def\bK{{\bf K}}
\def\bq{{\bf q}}
\def\br{{\bf r}}
\def\half{\frac{1}{2}}
\def\args{(\bx,t)}

\section{Introduction}\label{sec:intro} One of the major
challenges in the theoretical description of various spintronics
phenomena \cite{wolf2001}, such as current-induced magnetization
reversal \cite{slonczewski1996,berger1996,tsoi1998,myers1999} and
domain-wall motion
\cite{grollier2003,tsoi2003,yamaguchi2004,yamanouchi2004,klaui2005,beach2006,hayashi2007},
is their inherent nonequilibrium character. In addition to the
dynamics of the collective degree of freedom, the magnetization,
the nonequilibrium behavior manifests itself in the quasi-particle
degrees of freedom that are driven out of equilibrium by the
nonzero bias voltage. Due to this, the fluctuation-dissipation
theorem \cite{nicobook,LLstatphys} cannot be applied to the
quasi-particles. This, in part, has led to controversy surrounding
the theory of current-induced domain wall motion
\cite{tatara2004,barnes2005}.

Effective equations of motion for order-parameter dynamics that do
obey the equilibrium fluctuation-dissipation theorem often take
the form of Langevin equations, or their corresponding
Fokker-Planck equations \cite{nicobook,LLstatphys,riskenbook}. In
the context of spintronics the relevant equation is the stochastic
Landau-Lifschitz-Gilbert equation for the magnetization direction
\cite{brown1963,kubo1970,ettelaie1984,garciapalacios1998,smith2001,safonov2005,rossi2005}.
In this paper we derive the generalization of this equation to the
nonzero-current situation, for a simple microscopic model
consisting of a single spin coupled to two leads via an onsite
Kondo coupling. This model is intended as a toy-model for a
magnetic impurity in a tunnel junction
\cite{parcollet2002,bulaevskii2003,katsura2006}. Alternatively,
one may think of a nanomagnet consisting of a collection of spins
that are locked by strong exchange coupling. The use of this
simple model is primarily motivated by the fact that it enables us
to obtain analytical results. Because the microscopic starting
point for discussing more realistic situations has a similar form,
however, we believe that our main results apply qualitatively to
more complicated situations as well. Similar models have been used
previously to explicitly study the violation of the
fluctuation-dissipation relation \cite{mitra2005}, and the
voltage-dependence of the Gilbert damping parameter
\cite{katsura2006}. Starting from this model, we derive an
effective stochastic equation for the dynamics of the spin
direction using the functional-integral description of the
Keldysh-Kadanoff-Baym nonequilibrium theory \cite{stoof1999}. (For
similar approaches to spin and magnetization dynamics, see also
the work by Rebei and Simionato\cite{rebei2005}, Nussinov {\it et
al.} \cite{nussinov2005} and Duine {\it et al.} \cite{duine2007}.)
This formalism leads in a natural way to the path-integral
formulation of stochastic differential equations
\cite{duine2002,zinnjustinbook}. One of the attractive features of
this formalism is that dissipation and fluctuations enter the
theory separately. This allows us to calculate the strength of the
fluctuations even when the fluctuation-dissipation theorem is not
valid.

\begin{figure}
\vspace{-0.5cm} \centerline{\epsfig{figure=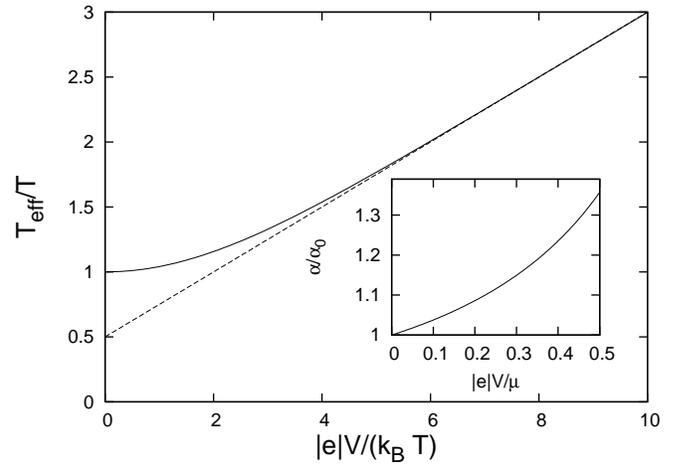}}
 \caption{Effective temperature as a function of bias voltage.
 The dashed line shows the large bias-voltage asymptotic result $k_{\rm B} T_{\rm eff} \simeq |e|V/4 + k_{\rm B} T/2$.
 The inset shows the bias-voltage dependence of the Gilbert damping parameter normalized to the zero-bias result.}
 \label{fig:teff}
\end{figure}

We find that the dynamics of the direction of the spin is
described by a Langevin equation with a damping kernel and a
stochastic magnetic field. We give explicit expressions for the
damping kernel and the correlation function of the stochastic
magnetic field that are valid in the entire frequency domain. In
general, they are not related by the fluctuation-dissipation
theorem. In the low-frequency limit the Langevin equation takes on
the form of the stochastic Landau-Lifschitz-Gilbert equation.
Moreover, in that limit it is always possible to introduce an
effective temperature that characterizes the fluctuations and the
equilibrium probability distribution for the spin direction. In
Fig.~\ref{fig:teff} we present our main results, namely the
bias-voltage dependence of the effective temperature and the
Gilbert damping parameter. We find that the Gilbert damping
constant initially varies linearly with the bias voltage, in
agreement with the result of Katsura {\it et al.}
\cite{katsura2006}. The voltage-dependence of the Gilbert damping
parameter is determined by the density of states evaluated at an
energy equal to the sum of the Fermi energy and the bias voltage.
The effective temperature is for small bias voltage equal to the
actual temperature, whereas for large bias voltage it is
independent of the temperature and proportional to the bias
voltage. This  bias-dependence of the effective temperature is
traced back to shot noise \cite{foros2005}.

\begin{figure}
\vspace{-0.5cm} \centerline{\epsfig{figure=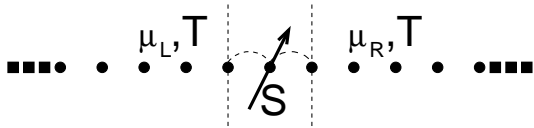,
width=7.0cm}}
 \caption{Model system of a spin $S$ connected to two tight-binding
 model half-infinite leads. The chemical potential
 of the left lead is $\mu_{\rm L}$ and different from the chemical potential of the right lead $\mu_{\rm R}$.
 The temperature $T$ of both leads is for simplicity taken to be equal.}
 \label{fig:system}
\end{figure}

Effective temperatures for magnetization dynamics have been
introduced before on phenomenological grounds in the context of
thermally-assisted current-driven magnetization reversal in
magnetic nanopillars \cite{urazhdhin2003,li2004,apalkov2005}. A
current-dependent effective temperature enters in the theoretical
description of these systems because the current effectively
lowers the energy barrier thermal fluctuations have to overcome.
In addition to this effect, the presence of nonzero current alters
the magnetization noise due to spin current shot noise
\cite{foros2005}. Covington {\it et al.} \cite{covington2004}
interpret their experiment in terms of current-dependent noise
although this interpretation is still under debate
\cite{rebei2005}. Foros {\it et al.} \cite{foros2005} also
predict, using a different model and different methods, a
crossover from thermal to shot-noise dominated magnetization noise
for increasing bias voltage. Our main result in
Fig.~\ref{fig:teff} is an explicit example of this crossover for a
specific model.

The remainder of the paper is organized as follows. We start in
Sec.~\ref{sec:langevin} by deriving the general Langevin equation
for the dynamics of the magnetic impurity coupled to two leads. In
Sec.~\ref{sec:equilibrium}~and~\ref{sec:nonequilibrium} we discuss
the low-frequency limit in the absence and presence of a current,
respectively. We end in Sec.~\ref{sec:concl} with our conclusions.

\section{Derivation of the Langevin equation} \label{sec:langevin}
We use a model that consists of a spin $S$ on a site that is
coupled via hopping to two semi-infinite leads, as shown in
Fig.~\ref{fig:system}. The full probability distribution for the
direction $\hat \Omega$ of the spin on the unit sphere is written
as a coherent-state path integral over all electron Grassmann
field evolutions $\psi^* (t)$ and $\psi (t)$, and unit-sphere
paths ${\bf S} (t)$, that evolve from $-\infty$ to $t$ and back on
the so-called Keldysh contour ${\mathcal C}^t$ . It is given by
\cite{stoof1999}
\begin{eqnarray}
\label{eq:probdistrfull}
  P[\hat \Omega,t] &=& \int_{{\bf S} (t) =  \hat \Omega}
   d [{\bf S}]~\delta\!\left[ \left|{\bf S}\right|^2-1\right] d[\psi^*]
  d[\psi] \nonumber \\
  &&\times \exp \left\{\frac{i}{\hbar} S[\psi^*,\psi,{\bf S}]
  \right\}~,
\end{eqnarray}
where the delta functional enforces the length constraint of the
spin. In the above functional integral an integration over
boundary conditions at $t=-\infty$, weighted by an appropriate
initial density matrix, is implicitly included in the measure. We
have not included boundary conditions on the electron fields,
because, as we shall see, the electron correlation functions that
enter the theory after integrating out the electrons are in
practice conveniently determined assuming that the electrons are
either in equilibrium or in the transport steady state.

The action $S [\psi^*,\psi,{\bf S}]$ is the sum of four parts,
\begin{eqnarray}
\label{eq:totalaction}
 S[\psi^*,\psi,{\bf S}]&=& S^{\rm L}\left[\left(\psi^{\rm L}\right)^*,\psi^{\rm
  L}\right]+S^{\rm R}\left[\left(\psi^{\rm R}\right)^*,\psi^{\rm
  R}\right]\nonumber \\
&+& S^{\rm C}\left[\left( \psi^0\right)^*,\psi^0,\left(\psi^{\rm
L}\right)^*,\psi^{\rm L},\left(\psi^{\rm R}\right)^*,\psi^{\rm
R}\right]
 \nonumber \\
&+& S^0 \left[\left( \psi^0\right)^*,\psi^0,{\bf S}\right]~.
\end{eqnarray}
We describe the leads using one-dimensional non-interacting
electron tight-binding models with the action
\begin{eqnarray}
\label{eq:actionleads}
 && S^{\rm L/R}\left[\left(\psi^{\rm L/R}\right)^*,\psi^{\rm L/R}\right] =
 \nonumber \\
&& \int_{{\mathcal C}^t} dt' \left\{
 \sum_{j,\sigma}
 \left(\psi^{\rm L/R}_{j,\sigma} (t') \right)^*  i \hbar \frac{\partial}{\partial t'}
 \psi^{\rm L/R}_{j,\sigma} (t') \right. \nonumber\\
 && \left.  + J \sum_{\langle j,j'\rangle;\sigma}
 \left(\psi_{j,\sigma}^{\rm L/R} (t')\right)^* \psi^{\rm L/R}_{j',\sigma}
 (t') \right\}~,
\end{eqnarray}
where the sum in the second term of this action is over nearest
neighbors only and proportional to the nearest-neighbor hopping
amplitude $J$ in the two leads. (Throughout this paper the
electron spin indices are denoted by $\sigma,\sigma'
\in\{\uparrow,\downarrow\}$, and the site indices by $j,j'$.) The
coupling between system and leads is determined by the action
\begin{eqnarray}
\label{eq:actioncoupling}
&&S^{\rm C}[\left(
\psi^0\right)^*,\psi^0,\left(\psi^{\rm L}\right)^*,\psi^{\rm
L},\left(\psi^{\rm R}\right)^*,\psi^{\rm R}] = \nonumber
\\ && \int_{{\mathcal C}^t}\!dt' J_{\rm C} \sum_{\sigma}\left[ \left( \psi^{\rm L}_{\partial {\rm
L},\sigma} (t') \right)^* \psi^0_{\sigma} (t')\!+\!
\left(\psi^0_{\sigma} (t') \right)^* \psi^{\rm L}_{\partial {\rm
L},\sigma} (t') \right]+
\nonumber \\
 && \int_{{\mathcal C}^t}\!dt' J_{\rm C}
\sum_{\sigma}\left[ \left( \psi^{\rm R}_{\partial {\rm R},\sigma}
(t') \right)^* \psi^0_{\sigma} (t') \!+\! \left(\psi^0_{\sigma}
(t') \right)^*
 \psi^{\rm R}_{\partial {\rm R},\sigma} (t')\right]~,\nonumber \\
\end{eqnarray}
where $\partial {\rm L}$  and $\partial {\rm R}$ denote the end
sites of the semi-infinite left and right lead, and the fields
$\left( \psi^0 (t) \right)^*$ and $ \psi^0 (t)$ describe the
electrons in the single-site system. The hopping amplitude between
the single-site system and the leads is denoted by $J_{\rm C}$.
Finally, the action for the system reads
\begin{eqnarray}
\label{eq:actionspin}
   && S^0\left[\left(\psi^0\right)^*,\psi^*,{\bf S}\right] =  \int_{{\mathcal C}^t} dt' \left[ \rule{0mm}{7mm}
  \sum_\sigma \left( \psi^0_\sigma (t') \right)^*
   i \hbar \frac{\partial}{\partial t'} \psi^0_\sigma (t')  \right. \nonumber\\ &&
  ~~~~~~~~~~-\hbar S {\bf A} ({\bf S} (t')) \cdot \frac{d {\bf S} (t')}{dt'}
   + {\bf h} \cdot {\bf S} (t') \nonumber
   \nonumber \\ && \left.
   ~~~~~~~~~~~~+\Delta \sum_{\sigma,\sigma'}
     \left( \psi^0_\sigma (t') \right)^* \bm{\tau}_{\sigma,\sigma'}
     \cdot {\bf S} (t')~\psi^0_{\sigma'} (t') \right]~.
\end{eqnarray}
The second term in this action is the usual Berry phase for spin
quantization \cite{auerbachbook}, with ${\bf A} ({\bf S})$ the
vector potential of a magnetic monopole
\begin{equation}
\label{eq:vectorpotmonopole}
  \epsilon_{\alpha\beta\gamma} \frac{\partial A_{\gamma}}{\partial S_{\beta}}
= S_\alpha~,
\end{equation}
where a sum over repeated Greek indices $\alpha,\beta,\gamma \in
\{x,y,z\}$ is implied throughout the paper, and
$\epsilon_{\alpha\beta\gamma}$ is the anti-symmetric Levi-Civita
tensor. The third term in the action in Eq.~(\ref{eq:actionspin})
describes the coupling of the spin to an external magnetic field,
up to dimensionful prefactors given by ${\bf h}$. (Note that {\bf
h} has the dimensions of energy.) The last term in the action
models the $s-d$ exchange coupling of the spin with the spin of
the conduction electrons in the single-site system and is
proportional to the exchange coupling constant $\Delta>0$. The
spin of the conduction electrons is represented by the vector of
the Pauli matrices that is denoted by $\bm{\tau}$.

Next, we proceed to integrate out the electrons using second-order
perturbation theory in $\Delta$. This results in an effective
action for the spin given by
\begin{eqnarray}
\label{eq:effactionspinfirst}
  && S^{\rm eff} [{\bf S}] = \int_{{\mathcal C}^t}
 dt' \left[ S
   \hbar {\bf A} ({\bf S} (t')) \cdot \frac{d {\bf S} (t')}{dt'}
   + {\bf h} \cdot {\bf S} (t') \right. \nonumber \\
   && \left. - \Delta^2 \int_{{\mathcal C}^t} dt''
     \Pi (t',t'') {\bf S} (t') \cdot {\bf S} (t'')
     \right]~.
\end{eqnarray}
This perturbation theory is valid as long as the electron band
width is much larger than the exchange interaction with the spin,
i.e.,  $J,J_{\rm C} \gg \Delta$. The Keldysh quasi-particle
response function is given in terms of the Keldysh Green's
functions by
\begin{equation}
\label{eq:qpresponsekeldysh}
  \Pi (t,t') = -\frac{i}{\hbar} G (t,t') G(t',t)~,
\end{equation}
where the Keldysh Green's function is defined by
\begin{equation}
\label{eq:gfkeldysh}
  i G (t,t') = \left\langle \psi^0_\uparrow (t) \left(\psi^0_\uparrow (t') \right)^*
  \right\rangle= \left\langle \psi^0_\downarrow (t) \left(\psi^0_\downarrow (t') \right)^*
  \right\rangle~.
\end{equation}
We will give explicit expressions for the response function and
the Green's function later on. For now, we will only make use of
the fact that a general function $A(t,t')$ with its arguments on
the Keldysh contour is decomposed into its analytic pieces by
means of
\begin{equation}
\label{eq:generaldecomp}
 A(t,t') = \theta (t,t') A^> (t,t') + \theta (t',t) A^< (t,t')~,
\end{equation}
where $\theta (t,t')$ is the Heaviside step function on the
Keldysh contour. There can be also a singular piece $A^{\delta}
\delta (t,t')$, but such a general decomposition is not needed
here. Also needed are the advanced and retarded components,
denoted respectively by the superscript $(-)$ and $(+)$, and
defined by
\begin{equation}
\label{eq:defretadvgen}
  A^{(\pm)}(t,t') \equiv \pm  \theta (\pm (t-t') ) \left[ A^> (t,t') - A^< (t,t')
  \right]~,
\end{equation}
and, finally, the Keldysh component
\begin{equation}
\label{eq:defkeldyshgen}
  A^{{\rm K}}(t,t') \equiv  A^> (t,t') + A^< (t,t')~,
\end{equation}
which, as we shall see, determines the strength of the
fluctuations.

Next we write the forward and backward paths of the spin on the
Keldysh contour, denoted respectively by ${\bf S} (t_+)$ and ${\bf
S} (t_-)$, as a classical path $\bm{\Omega} (t)$ plus fluctuations
$\delta \bm{\Omega} (t)$, by means of
\begin{equation}
\label{eq:splitfieldsS}
 {\bf S} (t_\pm) = \bm{\Omega} (t)  \pm \frac{\delta \bm{\Omega} (t)}{2}~.
\end{equation}
Moreover, it turns out to be convenient to write the delta
functional, which implements the length constraint of the spin, as
a path integral over a Lagrange multiplier $\Lambda (t)$ defined
on the Keldysh contour. Hence we have for the probability
distribution in first instance that
\begin{equation}
\label{eq:probdistrwithmultiplier}
  P [\hat \Omega,t]  =
    \int_{{\bf S} (t) =  \hat \Omega}  d[{\bf S}]
    d [\Lambda] \exp\left\{\frac{i}{\hbar} S^{\rm eff} [{\bf S}] +\frac{i}{\hbar} S^\Lambda [{\bf S},
    \Lambda]\right\}~,
\end{equation}
with
\begin{equation}
\label{eq:actionmuliplier}
  S^\Lambda [{\bf S},\Lambda] = \int_{{\mathcal C}^t} dt' \Lambda
  (t') \left [ \left| {\bf S} (t') \right|^2 -1  \right]~.
\end{equation}
We then also have to split the Lagrange multiplier into classical
and fluctuating parts according to
\begin{equation}
\label{eq:splitlagrange}
  \Lambda (t_\pm) = \lambda (t) \pm \frac{\delta \lambda (t)}{2}~.
\end{equation}
Note that the coordinate transformations in
Eqs.~(\ref{eq:splitfieldsS})~and~(\ref{eq:splitlagrange}) have a
Jacobian of one. Before we proceed, we note that in principle we
are required to expand the action up to all orders in $\delta
\bm{\Omega}$. Also note that for some forward and backward paths
${\bf S} (t_+)$ and ${\bf S} (t_-)$ on the unit sphere the
classical path $\bm{\Omega}$ is not necessarily on the unit
sphere. In order to circumvent these problems we note that the
Berry phase term in Eq.~(\ref{eq:actionspin}) is proportional to
the area on the unit sphere enclosed by the forward and backward
paths. Hence, in the semi-classical limit $S \to \infty$
\cite{katsura2006,auerbachbook} paths whose forward and backward
components differ substantially will be suppressed in the path
integral. Therefore, we take this limit from now on which allows
us to expand the action in terms of fluctuations $\delta
\bm{\Omega} (t)$ up to quadratic order. We will see that the
classical path $\bm{\Omega} (t)$ is now on the unit sphere. We
note that this semi-classical approximation is not related to the
second-order perturbation theory used to derive the effective
action.

Splitting the paths in classical and fluctuation parts gives for
the probability distribution
\begin{equation}
\label{eq:probdistrwithsplit}
  P [\hat \Omega,t]  =
    \int_{\bm{\Omega} (t) =  \hat \Omega}  \!\!\!\!\!\!\!\!\! d[\bm{\Omega}] d[\delta \bm{\Omega} ]
    d [\lambda] d[\delta \lambda] \exp\left\{\frac{i}{\hbar} S [\bm{\Omega}, \delta \bm{\Omega}, \lambda,\delta\lambda]\right\}~,
\end{equation}
with the action, that is now projected on the real-time axis,
\begin{eqnarray}
\label{eq:actionaftersplit}
 && S [\bm{\Omega}, \delta \bm{\Omega}, \lambda,\delta\lambda] =
 \int dt \left\{ \hbar S \epsilon_{\alpha\beta\gamma} \delta
 \Omega_\beta (t) \frac{d \Omega_\alpha (t)}{dt}
 \Omega_\gamma (t) \right. \nonumber \\
 &&\left.  +\delta\Omega_\alpha(t) h_\alpha
 + 2  \delta \Omega_\alpha (t) \Omega_\alpha (t)
 \lambda (t) \right.\nonumber \\
 && \left .  + \delta \lambda (t) \left[ |\bm{\Omega} (t) |^2-
 1+|\delta \bm{\Omega} (t)|^2/4\right]\rule{0mm}{5mm} \right\}
 \nonumber \\
 &- &\Delta^2 \int dt \int dt' \left\{ \delta \Omega_\alpha (t)
 \left[  \Pi^{(-)} (t',t)+\Pi^{(+)} (t,t') \right] \Omega_\alpha (t')\right\} \nonumber \\
 & -& \frac{\Delta^2}{2} \int dt \int dt' \left[  \delta \Omega_\alpha (t)
 \Pi^{\rm K} (t,t')\delta \Omega_\alpha (t')\right]~.
\end{eqnarray}
From this action we observe that the integration over $\delta
\lambda (t)$ immediately leads to the constraint
\begin{equation}
\label{eq:constraintonomega}
  \left| \bm{\Omega} (t) \right|^2=1- \frac{|\delta \bm{\Omega}
  (t)|^2}{4}~,
\end{equation}
as expected. Implementing this constraint leads to terms of order
${\mathcal O} (\delta \bm{\Omega}^3)$ or higher in the above
action which we are allowed to neglect because of the
semi-classical limit. From now on we can therefore take the path
integration over $\bm{\Omega} (t)$ on the unit sphere.

The physical meaning of the terms linear and quadratic in $\delta
\bm{\Omega} (t) $ becomes clear after a so-called
Hubbard-Stratonovich transformation which amounts to rewriting the
action that is quadratic in the fluctuations as a path integral
over an auxiliary field $\bm{\eta} (t)$. Performing this
transformation leads to
\begin{eqnarray}
\label{eq:probdistrwitheta}
  P [\hat \Omega,t] & =&
    \int_{\bm{\Omega} (t) =  \hat \Omega}   d[\bm{\Omega}] d[\delta \bm{\Omega}
    ]d[\bm{\eta}]
    d [\lambda]
    \nonumber \\
    && \times \exp\left\{\frac{i}{\hbar} S [\bm{\Omega},\delta \bm{\Omega},\lambda,\bm{\eta}]\right\}~,
\end{eqnarray} where the path integration over $\bm{\Omega}$ is now on the unit
sphere. The action that weighs these paths is given by
\begin{eqnarray}
\label{eq:actionaftersplitwithnoise}
 && S [\bm{\Omega}, \delta \bm{\Omega}, \lambda,\bm{\eta}] =
 \int dt \left[ \hbar S \epsilon_{\alpha\beta\gamma} \delta
 \Omega_\beta (t) \frac{d \Omega_\alpha (t)}{dt}
 \Omega_\gamma (t)  \right. \nonumber \\
 && \left. +\delta\Omega_\alpha(t) h_\alpha
 + 2  \delta \Omega_\alpha (t) \Omega_\alpha (t)
 \lambda (t) + \delta \Omega_\alpha (t) \eta_\alpha (t) \rule{0mm}{5mm} \right]
 \nonumber \\
 & -& \Delta^2 \int\!dt \int\! dt' \left\{ \delta \Omega_\alpha (t)
 \left[ \!\Pi^{(-)}\!(t',t)\!+\!\Pi^{(+)}\!(t,t')\!\right] \Omega_\alpha (t')\right\} \nonumber \\
 & +& \frac{1}{2 \Delta^2} \int dt \int dt' \left[  \eta_\alpha (t)
 \left(\Pi^{\rm K}\right)^{-1} (t,t') \eta_\alpha (t')\right]~.
\end{eqnarray}
Note that the inverse in the last term is defined as $\int dt''
\Pi^{\rm K} (t,t'') \left(\Pi^{\rm K}\right)^{-1} (t'',t') =
\delta (t-t')$.

Performing now the path integral over $\delta \bm{\Omega} (t)$, we
observe that the spin direction $\bm{\Omega} (t)$ is constraint to
obey the Langevin equation
\begin{eqnarray}
\label{eq:langevinfull}
  && \hbar S \epsilon_{\alpha\beta\gamma} \frac{d \Omega_\beta
  (t)}{dt}\Omega_\gamma (t) = h_\alpha + 2 \lambda (t)
  \Omega_\alpha (t) \nonumber \\ &&
  ~~~~~~~~~~~~~ + \eta_\alpha (t) + \int_{-\infty}^\infty dt' K (t,t')
  \Omega_\alpha (t')~,
\end{eqnarray}
with the so-called damping or friction kernel given by
\begin{equation}
\label{eq:frictionkernel}
  K (t,t') = -  \Delta^2 \left[ \Pi^{(-)} (t',t) + \Pi^{(+)} (t,t')
  \right]~.
\end{equation}
Note that the Heaviside step functions in
Eq.~(\ref{eq:defretadvgen}) appear precisely such that the
Langevin equation is causal. The stochastic magnetic field is seen
from Eq.~(\ref{eq:actionaftersplitwithnoise}) to have the
correlations
\begin{eqnarray}
\label{eq:noisecorrsgeneral}
  \langle \eta_\alpha (t) \rangle &=&0~; \nonumber \\
  \langle \eta_\alpha (t) \eta_\beta (t')\rangle &=&
  i \delta_{\alpha\beta} \hbar \Delta^2 \Pi^{\rm K} (t,t')~.
\end{eqnarray}
Using the fact that $\bm{\Omega} (t)$ is a unit vector within our
semi-classical approximation, the Langevin equation for the
direction of the spin $\hat \Omega (t)$ is written as
\begin{equation}
\label{eq:stochasticLLGfull}
  \hbar S \frac{d \hat \Omega(t)}{dt}
  = \hat\Omega (t) \bm{\times}
  \left[ {\bf h}  +  \bm{\eta} (t) + \int_{-\infty}^\infty dt' K (t,t')
  \hat\Omega (t') \right]~,
\end{equation}
which has the form of a Landau-Lifschitz equation with a
stochastic magnetic field and a damping kernel. In the next
sections we will see that for slowly-varying spin direction we get
the usual form of the Gilbert damping term.

So far, we have not given explicit expressions for the response
functions $\Pi^{(\pm),{\rm K}} (t,t')$. To determine these
functions, we assume that the left and right leads are in thermal
equilibrium at chemical potentials $\mu_{\rm L}$ and $\mu_{\rm
R}$, respectively. Although not necessary for our theoretical
approach we assume, for simplicity, that the temperature $T$ of
the two leads is the same. The Green's functions for the system
are then given by \cite{caroli1972,dattabook}
\begin{eqnarray}
\label{eq:lesserandgreatersystem}
 - i G^< (\epsilon) &=& \frac{A (\epsilon)}{2} \sum_{k \in \{\rm L,R\}}
 N(\epsilon-\mu_k)  ~; \nonumber \\
 i G^> (\epsilon) &=& \frac{A (\epsilon)}{2} \sum_{k \in \{\rm L,R\}}
 \left[ 1-N(\epsilon-\mu_k) \right] ~; \nonumber \\
 G^{\lessgtr,{\rm K}} (t-t') &=& \int \frac{d\epsilon}{(2\pi)} e^{-i \epsilon (t-t')/\hbar} G^{\lessgtr,{\rm K}} (\epsilon) ~,
\end{eqnarray}
with $N(\epsilon)=\{\exp[\epsilon/(k_{\rm B} T)]+1\}^{-1}$ the
Fermi-Dirac distribution function with $k_{\rm B}$ Boltzmann's
constant, and
\begin{equation}
\label{eq:spectrfct}
 A (\epsilon) = i\left[ G^{(+)} (\epsilon) - G^{(-)} (\epsilon) \right]~,
\end{equation}
the spectral function. Note that
Eq.~(\ref{eq:lesserandgreatersystem}) has a particularly simple
form because we are dealing with a single-site system. The
retarded and advanced Green's functions are determined by
\begin{equation}
\label{eq:eqnforgplusminus} \left[  \epsilon^\pm - 2 \hbar
\Sigma^{(\pm)} (\epsilon) \right] G^{(\pm)} (\epsilon) = 1~,
\end{equation}
with $\epsilon^\pm=\epsilon\pm i0$, and the retarded self-energy
due to one lead follows, for a one-dimensional tight-binding
model, as
\begin{equation}
\label{eq:selfenergylead}
  \hbar \Sigma^{(+)}(\epsilon) =
  -\frac{J_{\rm C}^2}{J}  e^{i k (\epsilon)a}~,
\end{equation}
with $k (\epsilon) =\arccos [-\epsilon/(2 J)]/a$ the wave vector
in the leads at energy $\epsilon$, and $a$ the lattice constant.
The advanced self-energy due to one lead is given by the complex
conjugate of the retarded one.

Before proceeding we give a brief physical description of the
above results. (More details can be found in
Refs.~[\onlinecite{caroli1972}]~and~[\onlinecite{dattabook}].)
They arise by adiabatically eliminating (``integrating out'') the
leads from the system, assuming that they are in equilibrium at
their respective chemical potentials. This procedure reduces the
problem to a single-site one, with self-energy corrections for the
on-site electron that describe the broadening of the on-site
spectral function from a delta function at the (bare) on-site
energy to the spectral function in Eq.~(\ref{eq:spectrfct}).
Moreover, the self-energy corrections also describe the
non-equilibrium occupation of the single site via
Eq.~(\ref{eq:lesserandgreatersystem})

For the transport steady-state we have that $\Pi^{(\pm),{\rm K}}
(t,t')$ depends only on the difference of the time arguments.
Using Eq.~(\ref{eq:qpresponsekeldysh}) and
Eqs.~(\ref{eq:generaldecomp}),~(\ref{eq:defretadvgen}),~and~(\ref{eq:defkeldyshgen})
we find that the Fourier transforms are given by
\begin{eqnarray}
\label{eq:responsefctfouriertrafoplusminus}
  &&\Pi^{(\pm)} (\epsilon) \equiv \int d(t-t') e^{i \epsilon
  (t-t')/\hbar}  \Pi^{(\pm)} (t,t') \nonumber \\
  &&=\int \frac{d\epsilon'}{(2\pi)} \int \frac{d\epsilon''}{(2\pi)}
  \frac{1}{\epsilon^\pm+\epsilon'-\epsilon''}  \nonumber \\
  &&~~~ \times \left[
    G^< (\epsilon') G^> (\epsilon'')-G^> (\epsilon') G^< (\epsilon'')\right]~,
\end{eqnarray}
and
\begin{eqnarray}
\label{eq:responsefctkeldyshcomponent} && \Pi^{\rm K} (\epsilon)
  = -2\pi i\int \frac{d\epsilon'}{(2\pi)} \int \frac{d\epsilon''}{(2\pi)}
  \delta (\epsilon+\epsilon'-\epsilon'')  \nonumber \\
  && ~~~ \times \left[  G^> (\epsilon') G^< (\epsilon'')
   + G^< (\epsilon')  G^> (\epsilon'')\right]~.
\end{eqnarray}
In the next two sections we determine the spin dynamics in the
low-frequency limit, using these expressions together with the
expressions for $G^\lessgtr (\epsilon)$. We consider first the
equilibrium case.

\section{Equilibrium Situation} \label{sec:equilibrium} In
equilibrium the chemical potentials of the two leads are equal so
that we have $\mu_{\rm L} = \mu_{\rm R} \equiv \mu$. Combining
results from the previous section, we find for the retarded and
advanced response functions (the subscript ``$0$" denotes
equilibrium quantities) that
\begin{eqnarray}
\label{eq:retadvresponseequilibrium}
  \Pi_0^{(\pm)} (\epsilon) &=&
   \int \frac{d\epsilon'}{(2\pi)} \int \frac{d\epsilon''}{(2\pi)} A(\epsilon') A(\epsilon'')
   \nonumber \\
   && \times
  \frac{\left[ N(\epsilon'-\mu) - N(\epsilon''-\mu) \right]}{\epsilon^\pm+\epsilon'-\epsilon''}~.
\end{eqnarray}
The Keldysh component of the response function is in equilibrium
given by
\begin{eqnarray}
\label{eq:keldyshcomponentrespequil}
    && \Pi_0^{\rm K} (\epsilon) = -2 \pi i
   \int \frac{d\epsilon'}{(2\pi)} \int \frac{d\epsilon''}{(2\pi)} A(\epsilon') A(\epsilon'') \delta(\epsilon-\epsilon'+\epsilon'')
   \nonumber \\
   && \!\!\!\!
  \left\{\left[1\!-\!N(\epsilon'\!-\!\mu)\right]N(\epsilon''\!-\!\mu) \!+\! N(\epsilon'\!-\!\mu)
   \left[1\!-\! N(\epsilon''\!-\!\mu)\right] \right\}.
\end{eqnarray}
The imaginary part of the retarded and advanced response functions
are related to the Keldysh component by means of
\begin{equation}
\label{eq:fdtheoremequil}
  \Pi_0^{\rm K} (\epsilon) = \pm 2 i \left[ 2 N_{\rm B} (\epsilon) +1 \right]
  {\rm Im} \left[ \Pi_0^{(\pm)} (\epsilon) \right],
\end{equation}
with $N_{\rm B} (\epsilon)=\{\exp[\epsilon/(k_{\rm B}
T)]-1\}^{-1}$ the Bose distribution function. This is, in fact,
the fluctuation-dissipation theorem which relates the dissipation,
determined as we shall see by the imaginary part of the retarded
and advanced components of the response function, to the strength
of the fluctuations, determined by the Keldysh component.

For low energies, corresponding to slow dynamics, we have that
\begin{eqnarray}
\label{eq:retadvresponseequillowenergy}
  \Pi_0^{(\pm)} (\epsilon) \simeq \Pi^{(\pm)}_0 (0)
       \mp \frac{i}{4\pi} A^2 (\mu) \epsilon~.
\end{eqnarray}
With this result the damping term in the Langevin equation in
Eq.~(\ref{eq:stochasticLLGfull}) becomes
\begin{equation}
\label{eq:dampingtermlowenergyequil}
   \int_{-\infty}^\infty dt' K (t,t')
  \hat\Omega (t') = - \frac{\hbar \Delta^2 A^2 (\mu)}{2 \pi} \frac{d \hat \Omega
  (t)}{d t}~,
\end{equation}
where we have not included the energy-independent part of
Eq.~(\ref{eq:retadvresponseequillowenergy}) because it does not
contribute to the equation of motion for $\hat \Omega (t)$. In the
low-energy limit the Keldysh component of the response function is
given by
\begin{equation}
\label{eq:responsekeldyshquillowenergy}
  \Pi^{\rm K}_0 (\epsilon) = \frac{A^2 (\mu)}{i \pi} k_{\rm B} T~.
\end{equation}

Putting all these results together we find that the dynamics of
the spin direction is, as long as the two leads are in equilibrium
at the same temperature and chemical potential, determined by the
stochastic Landau-Lifschitz-Gilbert equation
\begin{equation}
\label{eq:stochasticLLGequil}
  \hbar S \frac{d \hat \Omega(t)}{dt}
  = \hat\Omega (t) \bm{\times}
  \left[ {\bf h}  +  \bm{\eta} (t)  \right]
  -\hbar \alpha_0 \hat \Omega \bm{\times} \frac{d \hat \Omega (t)}{dt}~,
\end{equation}
with the equilibrium  Gilbert damping parameter
\begin{equation}
\label{eq:equilgilbertparameter}
 \alpha_0 =  \frac{\Delta^2 A^2 (\mu)}{2\pi}~.
\end{equation}
Using
Eqs.~(\ref{eq:noisecorrsgeneral}),~(\ref{eq:responsekeldyshquillowenergy}),~and~(\ref{eq:equilgilbertparameter})
we find that the strength of the Gaussian stochastic magnetic
field is determined by
\begin{equation}
\label{eq:noisecorrsequil} \langle \eta_\alpha (t) \eta_\beta (t')
\rangle =
 2   \alpha_0 \hbar k_{\rm B} T \delta (t-t') \delta_{\alpha\beta}~.
\end{equation}
Note that these delta-function type noise correlations are derived
by approximating the time dependence of $\Pi^{\rm K} (t,t')$ by a
delta function in the difference of the time variables. This means
that the noisy magnetic field $\bm{\eta} (t)$ corresponds to a
Stratonovich stochastic process
\cite{nicobook,LLstatphys,riskenbook}.

The stationary probability distribution function generated by the
Langevin equation in
Eqs.~(\ref{eq:stochasticLLGequil})~and~(\ref{eq:noisecorrsequil})
is given by the Boltzmann distribution
\cite{brown1963,kubo1970,ettelaie1984,garciapalacios1998,smith2001,safonov2005,rossi2005}
\begin{equation}
\label{eq:boltzmanndistrequil}
  P[\hat \Omega, t \to \infty] \propto \exp \left\{ - \frac{ E (\hat
  \Omega)}{k_{\rm B} T }\right\}~,
\end{equation}
with
\begin{equation}
\label{eq:energy}
   E [\hat \Omega] = - {\bf h} \cdot \hat \Omega~,
\end{equation}
the energy of the spin in the external field. It turns out that
Eq.~(\ref{eq:boltzmanndistrequil}) holds for any effective field
${\bf h} = -\partial E[\hat \Omega]/\partial \hat \Omega$, and in
particular for the case that $E[\hat \Omega]$ is quadratic in the
components of $\hat \Omega$ as is often used to model magnetic
anisotropy.

It is important to realize that the equilibrium probability
distribution has precisely this form because of the
fluctuation-dissipation theorem, which ensures that dissipation
and fluctuations cooperate to achieve thermal equilibrium
\cite{nicobook,LLstatphys}. Finally, it should be noted that this
derivation of the stochastic Landau-Lifschitz-Gilbert equation
from a microscopic starting point circumvents concerns regarding
the phenomenological form of damping and fluctuation-dissipation
theorem, which is subject of considerable debate
\cite{smith2001,safonov2005}.

\section{Nonzero bias voltage} \label{sec:nonequilibrium} In
this section we consider the situation that the chemical potential
of the left lead is given by $\mu_{\rm L} = \mu + |e|V$, with
$|e|V>0$ the bias voltage in units of energy, and $\mu=\mu_{\rm
R}$ the chemical potential of the right lead. Using the general
expressions given for the response functions derived in
Sec.~\ref{sec:langevin}, it is easy to see that the imaginary part
of the retarded and advanced components of the response functions
are no longer related to the Keldysh component by means of the
fluctuation-dissipation theorem in Eq.~(\ref{eq:fdtheoremequil}).
See also the work by Mitra and Millis \cite{mitra2005} for a
discussion of this point. As in the previous section, we proceed
to determine the low-frequency behavior of the response functions.

Using
Eqs.~(\ref{eq:lesserandgreatersystem}),~(\ref{eq:spectrfct}),~and~(\ref{eq:responsefctfouriertrafoplusminus})
we find that the retarded and advanced components of the response
function are given by
\begin{equation}
\label{eq:lowfrequencyplusminusneq}
  \Pi^{(\pm)} (\epsilon) = \mp \frac{i}{8 \pi} \left[ A^2 \left( \mu+ |e| V\right) + A^2 \left(\mu \right)
  \right] \epsilon~.
\end{equation}
In this expression we have omitted the energy-independent part and
the contribution following from the principal-value  part of the
energy integral because, as we have seen previously, these do not
contribute to the final equation of motion for the direction of
the spin. Following the same steps as in the previous section, we
find that the damping kernel in the general Langevin equation in
Eq.~(\ref{eq:stochasticLLGfull}) reduces to a Gilbert damping term
with a voltage-dependent damping parameter given by
\begin{eqnarray}
\label{eq:voltagedptgilbertdamping}
  \alpha \left(V \right)
   &=& \frac{\Delta^2 }{4 \pi}  \left[ A^2 \left( \mu+ |e| V\right) + A^2 \left(\mu \right)
  \right] \nonumber \\
  & \simeq& \alpha_0 \left[ 1+{ \mathcal O} \left( \frac{|e|V}{\mu}\right) \right]~.
\end{eqnarray}
This result is physically understood by noting that the Gilbert
damping is determined by the dissipative part of the response
function $\Pi^{(+)} (\epsilon)$. In this simple model, this
dissipative part gets contributions from processes that correspond
to an electron leaving or entering the system, to or from the
leads, respectively. The dissipative part is in general
proportional to the density of states at the Fermi energy. Since
the Fermi energy of left and right lead is equal to $\mu+|e|V$ and
$\mu$, respectively, the Gilbert damping has two respective
contributions corresponding to the two terms in
Eq.~(\ref{eq:voltagedptgilbertdamping}).

Note that the result that the Gilbert damping parameter initially
varies linearly with the voltage is in agreement with the results
of Katsura {\it et al.} \cite{katsura2006}, although these authors
consider a slightly different model. In the inset of
Fig.~\ref{fig:teff} we show the Gilbert damping parameter as a
function of voltage.  The parameters taken are $\Delta/J=0.1$,
$J_{\rm C}=J$, $\mu/J=1$ and $\mu/(k_{\rm B} T)=100$.

Although we can no longer make use of the fluctuation-dissipation
theorem, we are nevertheless able to determine the fluctuations by
calculating the low-energy behavior of the Keldysh component of
the response function in the nonzero-voltage situation. It is
given by
\begin{eqnarray}
\label{eq:keldyshcptresponsefctbias}
 &&  \Pi^{\rm K} (\epsilon) = - \frac{i}{2}
  \int \frac{d\epsilon'}{(2\pi)} A^2 (\epsilon')
  \left\{ \left[ N (\mu_{\rm L}\!-\!\epsilon')\!+\!N (\mu_{\rm R}\!-\!\epsilon')
  \right] \right. \nonumber\\
  && \left.  ~~~~~~~~~ \times
  \left[ N (\epsilon'\!-\!\mu_{\rm L})\!+\!N(\epsilon'\!-\!\mu_{\rm R})\right]
  \right\}~.
\end{eqnarray}
We define an effective temperature by means of
\begin{equation}
\label{eq:efftemp}
  k_{\rm B} T_{\rm eff} (T,V) \equiv
  \frac{i\Pi^{\rm K} (\epsilon) \Delta^2}{2 \alpha (V)}~.
\end{equation}
This definition is motivated by the fact that, as we mention
below, the spin direction obeys the stochastic
Landau-Lifschitz-Gilbert equation with voltage-dependent damping
and fluctuations characterized by the above effective temperature
\cite{arrachea2005}. From the expression for $\alpha (V)$ and
$\Pi^{\rm K} (\epsilon)$ we see that in the limit of zero bias
voltage we recover the equilibrium result $T_{\rm eff} = T$. In
the situation that $|e|V$ is substantially larger than $k_{\rm B}
T$, which is usually approached in experiments, we have that
\begin{equation}
\label{eq:efftemplargebias}
  k_{\rm B} T_{\rm eff} (T,V)  \simeq \frac{|e|V}{4} + \frac{k_{\rm B} T}{2}~,
\end{equation}
which in the limit that $|e|V \gg k_{\rm B } T$ becomes
independent of the actual temperature of the leads. In
Fig.~\ref{fig:teff} the effective temperature as a function of
bias voltage is shown, using the  expression for $\Pi^{\rm K }
(\epsilon)$ given in Eq.~(\ref{eq:keldyshcptresponsefctbias}). The
parameters are the same as before, i.e., $\Delta/J=0.1$, $J_{\rm
C}=J$, $\mu/J=1$ and $\mu/(k_{\rm B} T)=100$. Clearly the
effective temperature changes from $T_{\rm eff}=T$ at zero bias
voltage to the asymptotic expression in
Eq.~(\ref{eq:efftemplargebias}) shown by the dashed line in
Fig.~\ref{fig:teff}. The crossover between actual temperature and
voltage as a measure for the fluctuations is reminiscent of the
theory of shot noise in mesoscopic conductors \cite{dejong1997}.
This is not surprising, since in the single-site model we use the
noise in the equation of motion ultimately arises because of
fluctuations in the number of electrons in the single-site system,
and is therefore closely related to shot noise in the current
through the system. Foros {\it et al.} \cite{foros2005} calculate
the magnetization noise arising from spin current shot noise in
the limit that $|e|V \gg k_{\rm B} T$ and $|e|V \ll  k_{\rm B} T$.
In these limits our results are similar to theirs.

With the above definition of the effective temperature we find
that in the nonzero bias voltage situation the spin direction
obeys the stochastic Landau-Lifschitz-Gilbert equation, identical
in form to the equilibrium case in
Eqs.~(\ref{eq:stochasticLLGequil})~and~(\ref{eq:noisecorrsequil}),
with the Gilbert damping parameter and temperature replaced
according to
\begin{eqnarray}
\label{eq:replacements}
  \alpha_0 &\to& \alpha (V)~; \nonumber \\
  T &\to& T_{\rm eff} (T,V)~.
\end{eqnarray}
Moreover, the transport-steady-state probability distribution for
the direction of the spinis a Boltzmann distribution with the
effective temperature characterizing the fluctuations.

\section{Discussion and conclusions} \label{sec:concl}
We have presented a microscopic derivation of the stochastic
Landau-Lifschitz-Gilbert equation for a semi-classical single spin
under bias. We found that the Gilbert damping parameter is voltage
dependent and to lowest order acquires a correction linear in the
bias voltage, in agreement with a previous study for a slightly
different model \cite{katsura2006}. In addition, we have
calculated the strength of the fluctuations directly without using
the fluctuation-dissipation theorem and found that, in the
low-frequency regime, the fluctuations are characterized by a
voltage and temperature dependent effective temperature.

To arrive at these results we have performed a low frequency
expansion of the various correlation functions that enter the
theory. Such an approximation is valid as long as the dynamics is
much slower than the times set by the other energy scales in the
system such as temperature and the Fermi energy. Moreover, in
order for the leads to remain in equilibrium as the spin changes
direction, the processes in the leads that lead to equilibration
have to be much faster than the precession period of the
magnetizationspin. Both these criteria are satisfied in
experiments with magnetic materials. In principle however, the
full Langevin equation derived in Sec.~\ref{sec:langevin} also
describes dynamics beyond this low-frequency approximation. The
introduction of the effective temperature relies on the
low-frequency approximation though, and for arbitrary frequencies
such a temperature can no longer be uniquely defined
\cite{mitra2005}.

An effective temperature for magnetization dynamics has been
introduced before on phenomenological grounds
\cite{urazhdhin2003,li2004,apalkov2005}. Interestingly, the
phenomenological expression of Urazhdin {\it et al.}
\cite{urazhdhin2003}, found by experimentally studying thermal
activation of current-driven magnetization reversal in magnetic
trilayers, has the same form as our expression for the effective
temperature in the large bias-voltage limit
[Eq.~(\ref{eq:efftemplargebias})] that we derived microscopically.
Zhang and Li \cite{li2004}, and Apalkov and Visscher
\cite{apalkov2005}, have, on phenomenological grounds, also
introduced an effective temperature to study thermally-assisted
spin-transfer-torque-induced magnetization switching. In their
formulation, however, the effective temperature is proportional to
the real temperature because the current effectively modifies the
energy barrier for magnetization reversal.

Foros {\it et al.} \cite{foros2005} consider spin current shot
noise in the large bias-voltage limit and find for sufficiently
large voltage that the magnetization noise is dominated by shot
noise. Moreover, they also consider the low bias-voltage limit and
predict a crossover for thermal to shot-noise dominated
magnetization fluctuations. Our main result in Fig.~\ref{fig:teff}
provides an explicit example of this crossover for a simple model
system obtained by methods that are easily generalized to more
complicated models. In the experiments of Krivorotov {\it et al.}
\cite{krivorotov2004} the temperature dependence of the dwell time
of parallel and anti-parallel states of a current-driven spin
valve was measured. At low temperatures $k_{\rm B}T \lesssim |e|V$
the dwell times are no longer well-described by a constant
temperature, which could be a signature of the crossover from
thermal noise to spin current shot noise. However, Krivorotov {\it
et al.} interpret this effect as due to ohmic heating, which is
not taken into account in the model presented in this paper, nor
in the work by Foros {\it et al.} \cite{foros2005}. Moreover, in
realistic materials phonons provide an additional heat bath for
the magnetization, with an effective temperature that may depend
in a completely different manner on the bias voltage than the
electron heat-bath effective temperature. Nonetheless, we believe
that spin current shot noise may be observable in future
experiments and that it may become important for applications as
technological progress enables further miniaturization of magnetic
materials. Moreover, the formalism presented here is an important
step in understanding magnetization noise from a microscopic
viewpoint as its generalization to more complicated models is in
principle straightforward. Possible interesting generalizations
include making one of the leads ferromagnetic (see also
Ref.~[\onlinecite{fransson2006}]). Since spin transfer torques
will occur on the single spin as a spin-polarized current from the
lead interacts with the single-spin system, the resulting model
would be a toy model for microscopically studying the attenuation
of spin transfer torques and current-driven magnetization reversal
by shot noise. Another simple and useful generalization would be
enlarging the system to include more than one spin. The formalism
presented here would allow for a straightforward microscopic
calculation of Gilbert damping and adiabatic and nonadiabatic spin
transfer torques which are currently attracting a lot of interest
in the context of current-driven domain wall motion
\cite{grollier2003,tsoi2003,yamaguchi2004,yamanouchi2004,klaui2005,beach2006,hayashi2007}.
The application of our theory in the present paper is, in addition
to its intrinsic physical interest, chosen mainly because of the
feasibility of analytical results. The applications mentioned
above are more complicated and analytical results may be no longer
obtainable. In conclusion, we reserve extensions of the theory
presented here for future work.

It is a great pleasure to thank Allan MacDonald for helpful
conversations. This work was supported in part by the National
Science Foundation under grants DMR-0606489, DMR-0210383, and
PHY99-07949. ASN is partially funded by Proyecto Fondecyt de
Iniciacion 11070008 and  Proyecto Bicentenario de Ciencia y
Tecnolog\'ia, ACT027.

\end{document}